\documentclass[aps,prb,twocolumn,superscriptaddress,showpacs,showkeys,floatfix]{revtex4}

\usepackage{graphicx}
\usepackage{dcolumn}
\usepackage[pdftex]{color}
\definecolor{darkgreen}{rgb}{0.133,0.545,0.133}
\definecolor{orange}{rgb}{1.0,0.76,0.02}

\begin{document}

\title{As vacancies, Ga  antisites  and  Au impurities in Zincblende and Wurtzite GaAs nanowire segments from first principles}

\author{Yaojun A. Du}
\email[]{yaojun.du@uni-due.de}
\author{Sung Sakong}
\author{Peter Kratzer}
\affiliation{Fakult\"{a}t f\"{u}r Physik and Center for Nanointegration (CENIDE),  Lotharstra{\ss}e 1, 47048  Duisburg, Germany}


\begin{abstract}
In this paper some specific issues related to point defects in GaAs nanowires are addressed with the help of density functional theory calculations. These issues mainly arise from the growth of nanowires under conditions different from those used for thin films or bulk GaAs, such as the co-existence of zincblende and wurtzite polytypes, the use of gold particles as catalyst, and the arsenic-limited growth regime. 
Hence, we carry out density-functional calculations for As vacancies,  Ga$_{\rm As}$ antisites, and Au impurities in ZB and WZ GaAs crystals. 
Our results show that As vacancies can diffuse within in a ZB GaAs crystal with migration barriers of $\sim$1.9 eV.
Within  WZ GaAs, As vacancy diffusion is found to be anisotropic, with low barriers of 1.60 up to 1.79 eV (depending on doping conditions) in the $ab$-plane, while there are higher barriers of 2.07 to 2.44 eV to diffuse along the $c$-axis.
The formation energy of Au impurities is found to be generally much lower than those of arsenic vacancies or Ga$_{\rm As}$ antisites. 
Thus, Au impurities will be the dominant defects formed in Au-catalyzed nanowire growth.
Moreover, we find that it is energetically more favorable by 1 to 2 eV for an Au impurity to replace a lattice 
Ga atom than a lattice As atom in GaAs. An Au substitutional defect
for a lattice Ga atom in ZB GaAs is found to  create a charge transfer level 
in the lower half of the band gap. 
While our calculations locate this level  
at $E_{v} + 0.22$ eV,  taking into account the inaccuracy of the density functional that ought to be corrected by a downshift of $E_{v}$ by about 0.2~eV 
results in good agreement with the experimental result of $E_{v} + 0.4$ eV.
\end{abstract} 


\maketitle

\section{Introduction  \label{intro}}

Semiconductor nanowires have been emerging as a promising building
block for various nano-devices.~\cite{Samuelson2003} These
potential applications include light emitters,~\cite{Guichard2006nl}
solar cells,~\cite{Maiolo2007jacs} and
microelectronics.~\cite{Cui2001sci, Thelander2006} All these
applications require such a prerequisite that it is possible to keep the
concentration of intrinsic defects and impurities in the nanowires
below a given threshold so as to enable long exciton lifetimes and to
avoid unintentional doping due to electrically active impurities.  In
recent years, many experimental efforts~\cite{Jensen2004, Persson2004,
  Harmand2005, Dick2005, Dick2005NL} have been devoted to fast and
controllable growth of nanowires that is desired for reliable devices;
and a gold nanoparticle is often used as a catalyst to promote the
growth of GaAs nanowires. However, an Au droplet that accelerates an
underneath growing nanowire can also leave impurities within the GaAs
nanowire.~\cite{TambeNL10, breuerNL11} Previous experimental efforts
have focused on characterizing Au defect levels within a GaAs
crystal,~\cite{Hiesingerpssa76, Yanjes82, Pandianjjap91}.
Moreover, experimental studies employing X-ray energy-dispersive
spectroscopy~\cite{Persson2004, TambeNL10} indicate deviations from
stoichiometry in the nanowires close to the growth zone, and thus
point to the abundant intrinsic defects, in addition to Au impurities,
in GaAs nanowires.  While carrier lifetimes in core-shell nanowires
from Au-free self-catalyzed growth \cite{JabeenStPeter12,Bar-SadanNL12} were found
to be much longer than in nanowires from Au-catalyzed
growth\cite{breuerNL11}, they still fall short of the values reported
for bulk samples by several orders of magnitude.  This may indicate a
higher level of growth-related intrinsic defects (as compared to bulk)
even in Au-free nanowires.  

The focus of our interest is the defects that could be 
formed in nanowire growth below
the metal nanoparticle.
While the most abundant defect in low-temperature grown bulk GaAs is the As
antisite\cite{DabrowskiPRB40}, we expect this defect {\em not} to play
a role under the more Ga-rich growth conditions beneath a  metal
catalyst particle.  It is known that near the melting point and under the conditions typical of liquid-phase epitaxy, As vacancies are the dominant defects in GaAs.~\cite{Hurle} Furthermore, under
As-deficient conditions, the As lattice site might remain unoccupied
(As vacancy $V_{\rm As}$), it could be occupied by gallium (Ga
antisite Ga$_{\rm As}$), or, in case of growth with a gold catalyst,
may be occupied by an Au atom (Au$_{\rm As}$).  Since there is a
general interest in possibly detrimental effects of Au on the
properties of GaAs nanowires, we include 
Au$_{\rm Ga}$ defects into our study.

In this context, one 
has to consider that the effective growth conditions for
the nanowire material are probably very different from the moderately
arsenic-rich conditions commonly used in GaAs thin film growth. 
This is due to  the following reasons: In 
nanowire growth, material deposited from the vapor phase onto the substrate, the sidewalls and the 
metal nanoparticle can reach the 
interfacial area 
between the nanowire tip and the particle via surface diffusion~\cite{Haraguchi1997,
  Jensen2004, Dubrovskii2005, Pankoke2012} or diffusion through the liquid metal particle.~\cite{Kratzer2012} It is clear that sufficient Ga
atoms can reach the interfacial growth zone at the nanowire tip, since
the catalyst particle actually consists of a Au-Ga
alloy.~\cite{Harmand2005} 
The situation is less clear concerning the As supply, 
since the low-pressure solubility of As in gold is
low (in Au-catalyzed growth), or a high high temperature leads to As loss (in self-catalyzed growth).
In these cases, one would expect As-deficient growth conditions at the
nanowire--particle interface.  One possible scenario is that additional
arsenic can reach the growth zone via an As vacancy diffusion
mechanism through the GaAs nanowire.  Moreover, under As-deficient conditions,
one would also expect that As vacancies could be left behind within the 
GaAs nanowire by the advancing interfacial growth zone. A subsequent
annealing process in arsenic vapor might be helpful to remove these
vacancies. Therefore, it is important to understand 
$V_{\rm As}$ diffusion within GaAs bulk, in order to comprehensively
understand the As supply path during the growth process and to
estimate a suitable annealing temperature for removing $V_{\rm As}$s
in a nanowire. A previous theoretical study based on density
functional theory (DFT)~\cite{hohenberg:1964,kohn:1965} calculated
relatively high migration barriers of $\sim 2.4$ eV for a $V_{\rm As}$
in zincblende GaAs.~\cite{Mousseau2007apa} This seemingly implies that
As transport through the nanowire could be inefficient.  However, the 
convergence of the atomic-orbital basis sets employed in these DFT
calculations has not been demonstrated.

While wurtzite GaAs cannot be obtained in bulk form by ordinary growth
techniques, the GaAs nanowires grown with Au catalyst may exhibit both
zincblende (ZB) and wurtzite (WZ) structures, or alternating segments
of these (and other) polytypes through stacking
faults.\cite{Persson2004, GerthsenPSS05, SpirkoskaPRB09,
  riechertPRB12} Hence, it is crucial to understand $V_{\rm As}$,
Ga$_{\rm As}$, and Au defects in {\em both} ZB and WZ GaAs crystals.
There is considerable knowledge, both from experimental\cite{Hurle}
and theoretical \cite{Schulz,Komsa} sides, about intrinsic point
defects in ZB GaAs.  However, experimental data on point defects in WZ
GaAs is still elusive.  In this work, we will perform plane-wave DFT
calculations to study various properties associated with $V_{\rm As}$,
Ga$_{\rm As}$, and Au defects in ZB and WZ GaAs crystals, and to
investigate and compare $V_{\rm As}$ diffusion in both GaAs
polytypes. This paper is organized as follows: The computational
approach and supercell models are described in Sec.~\ref{model}.  The
formation energies of $V_{\rm As}$, Ga$_{\rm As}$ defects and Au
impurities in ZB and WZ GaAs crystals are discussed in
Sec.~\ref{AuGaAs}.  The diffusion of $V_{\rm As}$s in various charge
states in ZB and WZ GaAs crystals are presented in Sec.~\ref{VGaAs}.
We summarize our work in Sec.~\ref{dis}.

\section{Methods of Calculation  \label{model}}

\begin{figure}
\includegraphics[width=0.4\textwidth]{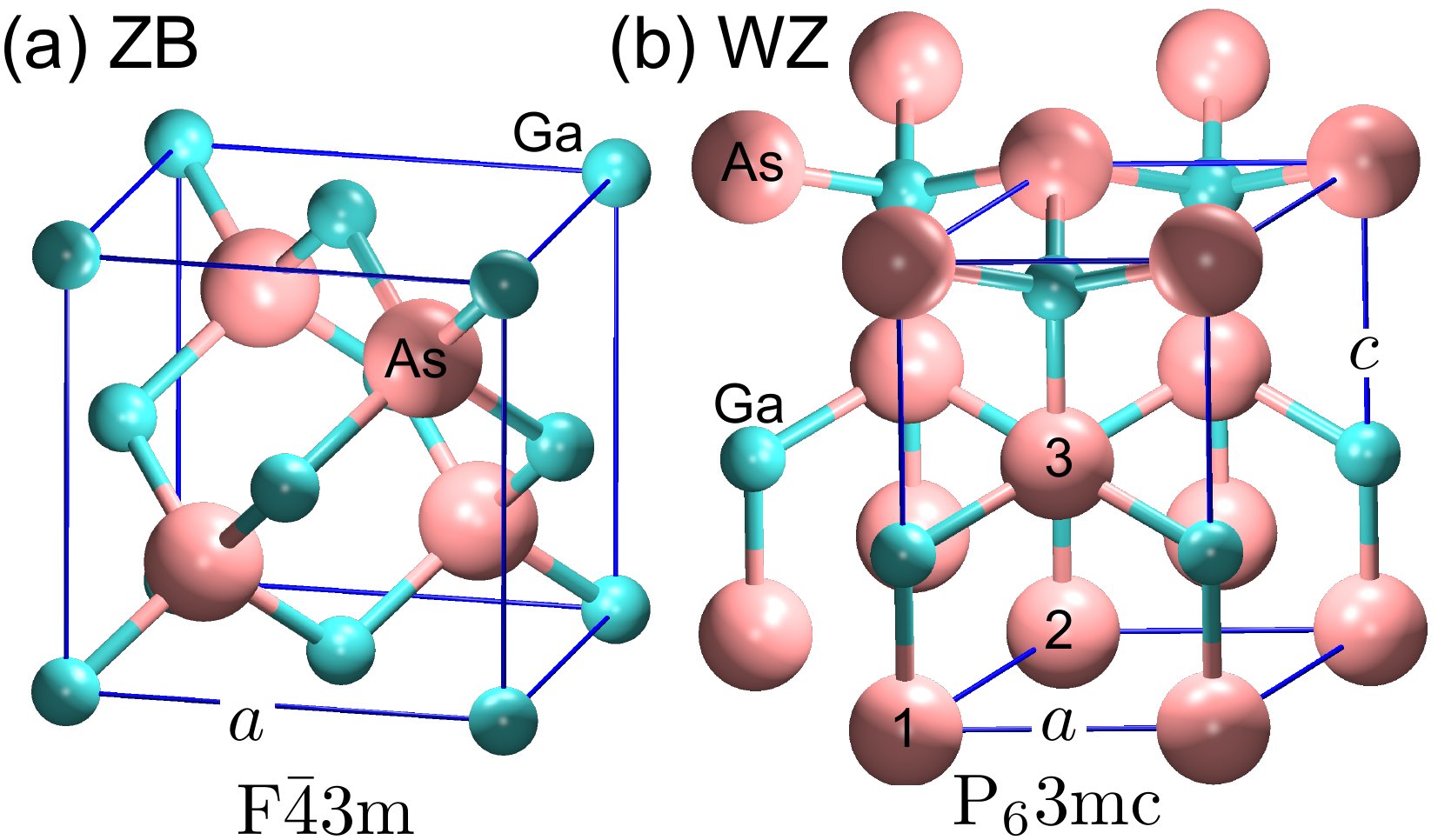}
\caption{(Color online)  The structure of ZB and WZ GaAs crystals. The conventional unit cell
for a ZB and a  WZ GaAs crystal is shown in (a) and (b), respectively. The As and Ga atoms 
are represented by large pink and small green spheres, respectively. 
\label{fig:crystals}}
\end{figure}

In this work, we have performed DFT calculations to study $V_{\rm
  As}$, Ga$_{\rm As}$ and substitutional Au impurities in ZB and WZ
GaAs using the supercell method. The conventional unit cell for a ZB
and a WZ GaAs crystal is shown in
Fig.~\ref{fig:crystals}. Sufficiently large supercells of 216 atoms
(ZB) or 96 atoms (WZ) are used to fully relax the strain induced by
defects and impurities.  We employ the projector augmented-wave (PAW)
method~\cite{blochl:1994, kresse:1999} as implemented in the VASP
code~\cite{vasp1:1993, vasp2:1996} for all DFT calculations.  The PAW
potentials include $4s4p$, $4s4p$, and $5d6s$ electrons as valence
electrons for As, Ga, and Au, respectively.  The generalized-gradient
approximation (GGA) is used for the exchange-correlation
functional.~\cite{perdew:1996} The Brillouin zone integration is
performed using the Monkhorst-Pack~\cite{monkhorst:1976} scheme over a
$(2\times 2 \times 2)$ k-point mesh 
(shifted with respect to the $\Gamma$-point)  
for both ZB and WZ
supercells. Then, the partition length is 0.06 Bohr$^{-1}$ or smaller
in each direction.  Spin-polarized calculations are performed for
supercells that contain an odd number of electrons. 
The optimized kinetic energy
cutoff of $E_{\rm cut } = 250$~eV  is used to compute the formation 
energies of various impurities and defects, whereas an  $E_{\rm cut}$ of 
200 eV is used for studying  $V_{\rm As}$ diffusion in GaAs.
 The optimized parameters allow us to
obtain a relative energy convergence of within 10 meV for all systems
presented in this study.  The method of a homogeneous background
charge is used to model charge states of the defect within the
supercell approach.  We determined various diffusion paths for  $V_{\rm As}$s
 within the As sublattice of
 both ZB and WZ GaAs crystals, employing the nudged-elastic band (NEB)
method.~\cite{jonsson:1998,jonsson:2000,jonsson:2000a} Here, the
migration pathway is represented by several intermediate
configurations 
between two fully relaxed end points and each image is relaxed until
the perpendicular forces with respect to the minimum energy path are
less than a given tolerance, which is set to be 0.03 eV/\AA\ in our
calculations.

For ZB GaAs (space group F$\bar{4}$3m), 
we have obtained the optimized lattice constant of $a = 5.76$ \AA, in good agreement  
with the experimental value of $a = 5.653$  \AA.~\cite{DST} 
For WZ GaAs  (space group P$_6$3mc), 
we have computed  the lattice constant to be $a = 4.05$ \AA\ and  $c = 6.69$ \AA, which compares well with 
previous GGA results of  $a = 4.050$ \AA\ and  $c = 6.678$ \AA.~\cite{walterPRB11}
The calculated direct band gaps of GaAs are 0.17~eV and 0.22 ~eV in ZB and WZ crystals, respectively. 
These results reflect the well-known underestimation of the electronic band gap by semi-local DFT functionals, in particular, when the (too large) theoretical lattice constant is used. 
In addition, we have performed DFT calculations with the same geometries as in GGA, but using the hybrid functional 
proposed by Heyd, Scuseria, and Ernzerhof (HSE).\cite{Heyd2003}  Our calculations include  25\% of the  exact exchange, and 
 the screening parameter is set to be 0.2 \AA$^{-1}$.
The band gaps, that are computed to be  1.15~eV and 0.99~eV in ZB and WZ GaAs, respectively, are much closer to experimental values. 
Moreover, we conclude from these calculations that the valence band maximum experiences a down-shift of 0.48~eV in ZB GaAs (0.36~eV in WZ GaAs). These values are in good agreement with recent literature~\cite{Chen&Pasquarello2012}, and will be used as corrections when
our computed results are compared to experiment.
The orthorhombic Ga (space group Cmca) crystal and 
trigonal As (space group R$\bar{3}$m) crystal are used to compute the Ga and As chemical potentials.

Since ZB and WZ segments co-exist in GaAs nanowires, the band edges
from ZB and WZ supercell calculations should be aligned
accordingly. We have constructed a 50 atom supercell of  a heterostructure of WZ and ZB GaAs.
The supercell  consists of 13 ZB and 12 
 WZ bilayers that stack along $[\bar{1}\bar{1}\bar{1}]$ or [0001] for a ZB
or a WZ segment, respectively.  The constructed heterostructure is optimized
for both the lattice constant and internal coordinates. 
Note that the lateral dimension of the  heterostructure happens to 
coincide with the average of the WZ and ZB lattice parameters.
The valence band offset of the ZB and WZ segments can be computed~\cite{Franciosi1996} as 
\begin{eqnarray}
\Delta E_{VB}({\rm GaAs}) &=& E_{v}^{\rm ZB} - E_{v}^{\rm WZ} - (E_{core}^{\rm ZB} ({\rm As}) - E_{core}^{\rm WZ} ({\rm As}) ) \nonumber \\
& & +(E_{core}^{\rm ZB~seg} ({\rm As}) - E_{core}^{\rm WZ~seg} ({\rm As}) ),
\label{offset}
\end{eqnarray}
where $E_{v}^{\rm ZB}$ and $E_{v}^{\rm WZ}$ are the valence band top energies at the $\Gamma$-point for ZB 
and WZ GaAs, respectively, $E_{core}^{\rm ZB} ({\rm As})$ and $E_{core}^{\rm WZ} ({\rm As})$ are the 
core level energies of As atoms in bulk ZB and WZ GaAs, respectively,
$E_{core}^{\rm ZB~seg} ({\rm As})$ and $E_{core}^{\rm WZ~seg} ({\rm As})$ are the 
core level energies of As atoms in ZB and WZ segments of the  heterostructure, respectively.
We compute the band offset to be  $\Delta E_{VB}({\rm GaAs})  = -0.0601$ eV, in good agreement 
with the value of $-0.0632$ eV from a previous calculation.~\cite{De2010}

\section{The energetics of defects and A\lowercase{u} impurities in G\lowercase{a}A\lowercase{s} Nanowires  \label{AuGaAs}}

\begin{figure}
\includegraphics[width=7cm]{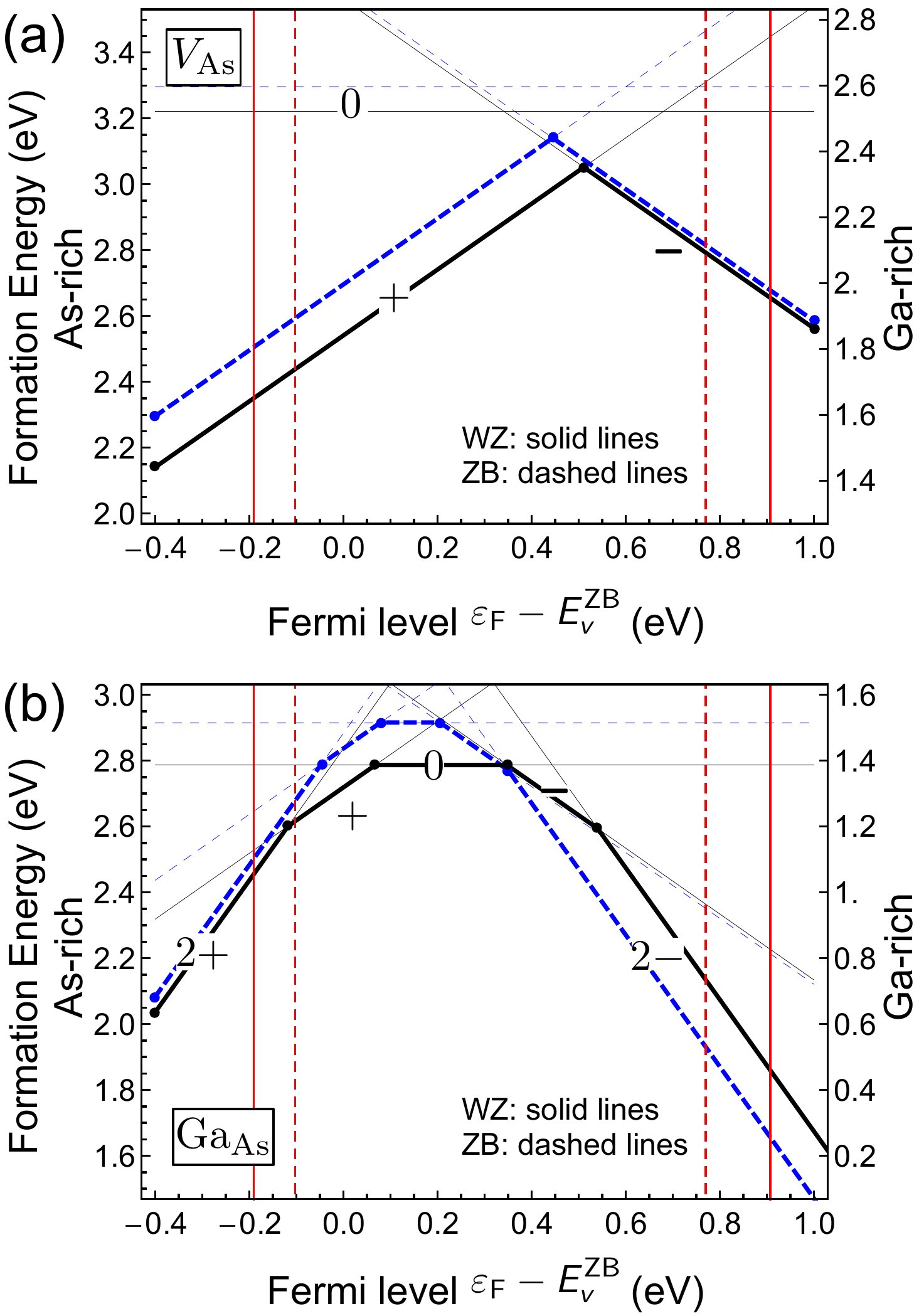}
\caption{(Color online)  The formation energy of an As vacancy (a) and a Ga antisite (b) in various charge states
 in GaAs polytypes at different doping conditions, as a function of the Fermi level $\varepsilon_\mathrm{F}$.   
The formation energies of   As vacancies in ZB GaAs and  WZ GaAs are shown as dashed and solid lines, respectively. 
The thick lines represent stable charge states.
The vertical lines indicate the HOMO and LUMO levels in the defect-free supercell and thus delimit the energy interval where 
the GGA total energies are considered reliable.
The left and right $y$-axes show the  formation energies under As-rich ($\mu_\mathrm{As}^{upper}$) and Ga-rich ($\mu_\mathrm{As}^{lower}$)
 conditions as described in the text.
Here, one has the identity  $\mu_{\rm As}^{lower} =  \mu_{\rm As}^{upper} - 0.7~{\rm eV}$.
\label{fig:EformVZBGaAs}}
\end{figure}

We start by calculating the formation energies of  relevant intrinsic defects. The formation energy 
of  a $q$-charged ($q$ is an integer number) defect (or impurity) $X$ at site $Y$ as a
 function of the Fermi level $\varepsilon_\mathrm{F}$ is defined by 
\begin{equation}
E_f \left[ X_Y^q, \varepsilon_\mathrm{F} \right]  = E \left[X_{Y}^q \right] - E_{host} 
-  \sum_i n_i\mu_i + q( \varepsilon_\mathrm{F} + E_{v}) + E_{corr} .  
  \label{efmvatas}
\end{equation}
Here, $E \left[X_Y^q \right]$ is the total energy of a $q$-charged supercell with a defect $X_Y$ 
 and $E_{host}$ is the energy of a perfect GaAs bulk supercell with the same shape. 
$n_i$ and $\mu_i$ represent the change in the number of species $i$ in the supercell and the chemical 
potential for a species $i$, respectively.  $\varepsilon_\mathrm{F}$ is the Fermi level that depends on the doping 
condition, and $E_{v}$ is the  valence band top energy of the corresponding bulk GaAs crystal.
Due to the $(2 \times 2 \times 2)$ k-point mesh used for supercell calculations, the gap between the highest occupied (HOMO) and the lowest unoccupied (LUMO) orbitals  is much higher in these supercell calculations compared to the gap at the $\Gamma$-point in GaAs bulk calculations. Even though this widened gap is a technical artifact of our method, it puts us in position to vary $\varepsilon_\mathrm{F}$ in a rather wide range while still maintaining the physically correct charge distribution localized near the defect. The energy interval around $E_{v}$ in which our calculations can be expected to yield physically meaningful charge transfer levels, as a charged defect cannot transfer charge to the LUMO or from the HOMO, is indicated in the figures. 
The linear and quadratic correction term with respect to $q$ of  $E_{corr}=\Delta V q + \alpha q^2/2 \epsilon L$  
accounts for the spurious electrostatic interaction between periodic images of charged defects~\cite{VandeWalle2004,MakovPRB51}. 
$\Delta V$ is a correction for the spurious potential off-set induced by the finite defect concentration in  the calculation; 
it is calculated from the  energy difference of atomic core levels 
between a neutral supercell with an   $X_Y$  defect and a perfect bulk supercell.
The values for $\alpha$ and $\epsilon$, the Madelung constant  
and the static dielectric constant, respectively, are taken according to the ZB or WZ polytype of bulk GaAs, 
and $L$ is the  supercell dimension  used in the corresponding calculations.
The values of    $\alpha$, $\epsilon$, and $L$ are listed in Table~\ref{vpara} in the Appendix for
the cases of ZB and WZ GaAs crystals, respectively. 
Our computed dielectric constant $\epsilon$ for ZB GaAs is 32\% larger than the experimental
 value of  0.90   $e$\AA$^{-1}$V$^{-1}$.~\cite{Moore1996} 
At the growth condition, the chemical potentials of Ga and As are in equilibrium with bulk GaAs. 
Thus, one has the identity  $\mu_\mathrm{Ga}+\mu_\mathrm{As}=\mu_\mathrm{GaAs}$, where 
$\mu_\mathrm{GaAs}=-E_\mathrm{GaAs}^{bulk}$ is  the cohesive energy per formula unit of a pertinent GaAs polytype.
The range of the As chemical potential is specified in accordance with the growth condition, with  the upper bound 
set to be  at the equilibrium with bulk As, i.e., $\mu_{\rm As}^{upper} =  -E_{\rm As}^{bulk}$.
The lower bound is set to be  at the equilibrium with bulk Ga as $\mu_{\rm As}^{lower} =  -E_{\rm GaAs}^{bulk}+E_{\rm Ga}^{bulk}$.
Disregarding the difference of the cohesive 
energies of GaAs of  20~meV  in ZB and WZ crystals, we align the upper bound $\mu_{\rm As}^{upper}$ for ZB and WZ to  
the same value. 
We take the lower bound  $\mu_{\rm As}^{lower}$ to be 0.7 eV below $\mu_{\rm As}^{upper}$, so that 
$\mu_{\rm As}^{lower}$ is very close to a Ga-rich condition for both ZB and WZ GaAs. The bulk energies of GaAs, Ga and As crystals  are calculated using the optimized lattice parameters.
Note that the upper and lower bounds of the As chemical potentials
correspond to As- and Ga-rich growth conditions, respectively.
In addition, the 
cohesive energy bulk Au per atom is inserted for the Au chemical potential, i.e., $\mu_\mathrm{Au}=-E_\mathrm{Au}^{bulk}$.

Figure~\ref{fig:EformVZBGaAs}(a) shows the  $V_\mathrm{As}$  formation energies for various charge states $q$, indicating 
the formation energies of $V_\mathrm{As}$ in WZ are overall lower than those in ZB.
The formation energies are computed using  Eq.~\ref{efmvatas}, and the potential off-sets $\Delta V$ associated with the $V_\mathrm{As}$s 
are listed in Table~\ref{vaupara} in the Appendix.   
Under an As-rich growth condition (left $y$-axis in Fig.~\ref{fig:EformVZBGaAs}(a)), the formation energies of  neutral $V_\mathrm{As}$s are 
quite high (3.3 eV and 3.2 eV in ZB and WZ, respectively).
Under a Ga-rich condition  (right $y$-axis in Fig.~\ref{fig:EformVZBGaAs}(a)), 
the formation energies are lowered by 0.7 eV; 
however, the energies remain  relatively high. 
Thus, the formation of As vacancies  could be energetically unfavorable  
under usual GaAs nanowire growth conditions in both ZB and WZ segments. 
Our results show that $V_{\rm As}^{+}$ and  $V_{\rm As}^{-}$ defects are  stable at various doping conditions in ZB and WZ GaAs crystals.
Deep $(+/-)$ levels are found at $E_v^\mathrm{ZB}+0.44$ and $E_v^\mathrm{WZ}+0.45$~eV in ZB 
and WZ, respectively. Since 
GaAs nanowires may have coexisting ZB and WZ segments, it is instructive to specify 
the charge transfer levels of defects in the  WZ segments also with respect to the valence band edge of the ZB segments, using our results that 
the valence band maximum of ZB GaAs is $0.0601$ eV lower than in WZ GaAs.  
Thus, one obtains the $(+/-)$ level in WZ to be $E_v^\mathrm{ZB}+0.51$~eV. 
We found that $V_{\rm As}^{2-}$ and  $V_{\rm As}^{3-}$ in both ZB and WZ GaAs crystals are unstable, consistent with Ref.~\onlinecite{Schulz}, 
because the structures relax to a GaAs antisite that is neighboring a Ga vacancy. 
As seen from  Fig.~\ref{fig:EformVZBGaAs}, the neutral As vacancies are energetically unfavorable, 
indicating that a $V_{\rm As}$ in GaAs is a so-called 'negative $U$' system. 
Our results are consistent with a previous theoretical studies~\cite{Mousseau2005prb, Schulz,Komsa2012}, although the quoted absolute  positions of  the $(+/-)$ levels in these studies 
are different from ours. 
For a quantitative comparison, it is important to realize that Schulz and von Lilienfeld~\cite{Schulz} attempted to set $E_{v}$ to the true ionization potential, whereas we use the plain DFT-GGA value. Estimating the true valence band top from our HSE calculations, a down-shift of $~0.4$~eV (see Section~\ref{model}) should be applied to our $E_{v}$ value in order to compare to their work, or to experimental data. The same applies when comparing to the values of Komsa and Pasquarello,~\cite{Komsa2012} who performed all their calculations with the HSE functional.
With this in mind, our results are in much better agreement with these previous works.
It is worth noting  that the negative $U$ feature of a $V_\mathrm{As}$ disappears in their HSE calculations. 
This is due to the fact that  the 
(negative) contribution of the electronic exchange energy, which is more pronounced in the hybrid functional, energetically favors spin-polarized solutions, e.g.,  the neutral $V_{\rm As}$  in the present case. 
Whether this stabilization of a $V_{\rm As}^0$ in a narrow range of the Fermi level is indeed a physical feature or not 
should be determined by future experiments.

The Ga antisite Ga$_{\rm As}$ could be an another intrinsic defect within a growing GaAs nanowire under  As-deficient 
conditions, and we show  the formation energies of Ga$_{\rm As}$ as a function of  $\varepsilon_\mathrm{F}$ 
in Fig.~\ref{fig:EformVZBGaAs}(b).
The formation energies are computed using Eq.~\ref{efmvatas}, with the potential off-set $\Delta V$ associated with Ga$_\mathrm{As}$ 
 defects listed in Table~\ref{vaupara} in the Appendix.
The formation energies are specified for both  Ga-rich (right $y$-axis in Fig.~\ref{fig:EformVZBGaAs}(b))  
and As-rich (left $y$-axis Fig.~\ref{fig:EformVZBGaAs}(b)) conditions, and 
formation energies are 1.4 eV lower in a Ga-rich condition than in an As-rich condition.  
In both ZB and WZ GaAs crystals, Ga$_{\rm As}$s with charge states from $2+$ to $2-$  may be stable under certain doping 
conditions. The associated formation energies in
 WZ GaAs are overall lower than in ZB GaAs. 
The Ga$_{\rm As}^{0}$ and Ga$_{\rm As}^{+}$ defects are energetically more favorable by 0.13-0.14 eV  in WZ than in ZB, whereas a  Ga$_{\rm As}^{-}$ defect  is 
only 0.01 eV more stable in WZ. As a result, the charge transfer levels $(0/-)$ differ by 0.12 eV  in ZB and WZ.
It follows that, on an absolute scale, the charge 
transfer levels are  located at $E_v^\mathrm{ZB}+0.21$ and 0.33 eV in ZB and WZ segments of a GaAs nanowire, respectively. We notice that both intrinsic 
defects of $V_{\rm As}$ and Ga$_{\rm As}$ have deep charge transfer levels and they are deeper in WZ GaAs. 
Comparing Fig.~\ref{fig:EformVZBGaAs}(a) and (b), one may find that 
$V_{\rm As}$s are more stable than Ga$_{\rm As}$s for the doping conditions of $\varepsilon_\mathrm{F}
-E_v^\mathrm{ZB} <0.21$ (0.27) eV in ZB (WZ)
under As rich conditions, 
while Ga$_{\rm As}$s are more stable than $V_{\rm As}$s for the doping conditions 
of $\varepsilon_\mathrm{F} -E_v^\mathrm{ZB} >0.21$ (0.27) eV in ZB (WZ). 
When comparing to experiment, one should take into account that the ``true" valence band top is lower than the $E_v$
 resulting from our DFT-GGA calculations, such that the range of horizontal-axis values plotted 
in Fig.~\ref{fig:EformVZBGaAs} approximately reflects the ``true'' band gap.

\begin{figure}
\includegraphics[width=7cm]{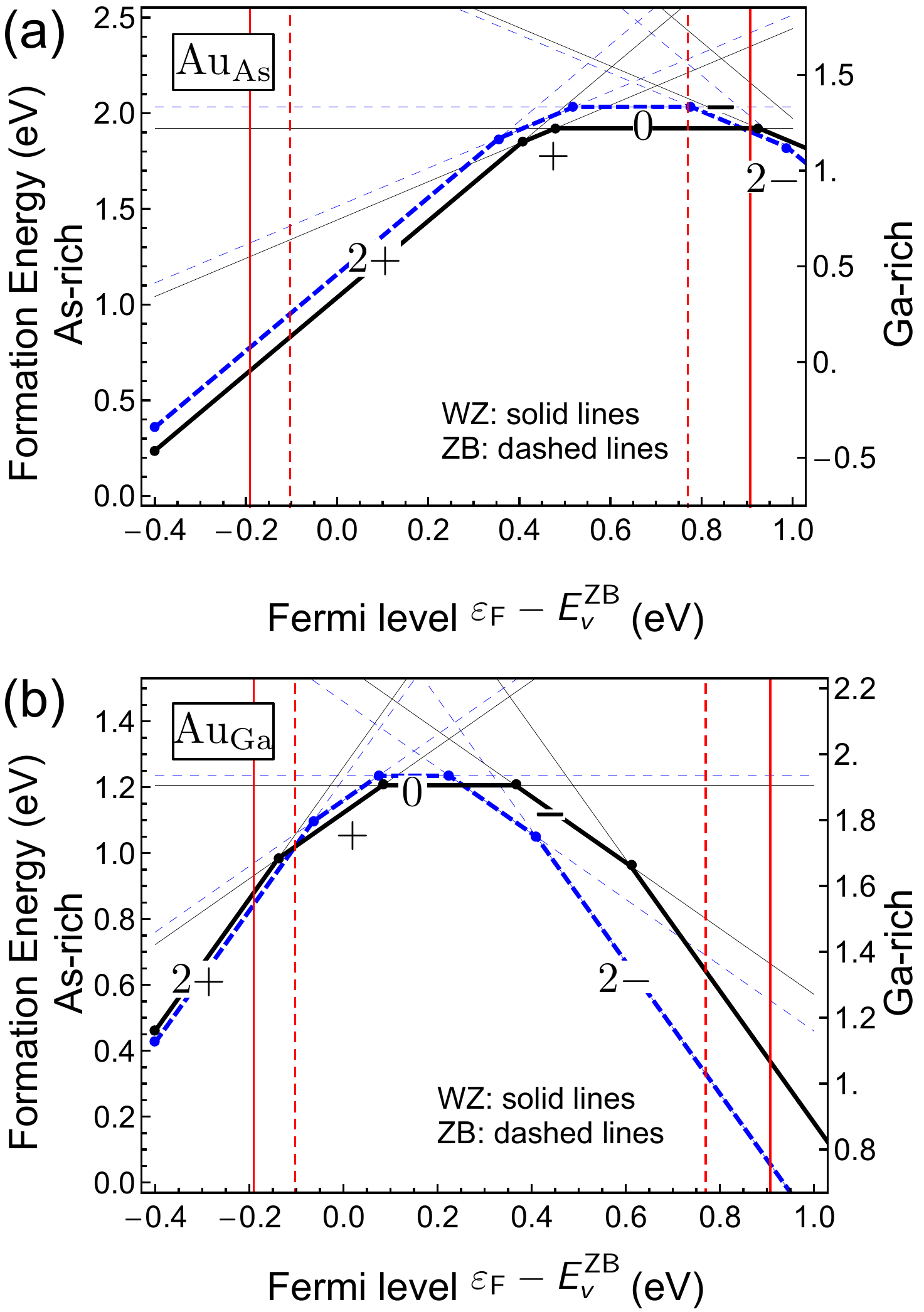}
\caption{(Color online)  The formation energies of a substitutional Au defect in the $\pm$1-charged, 
 $\pm$2-charged, and neutral states in WZ and ZB GaAs crystals. The formation energies of a substitutional Au impurity
for a lattice As atom and a lattice Ga atom are shown in (a) and (b), respectively.
The formation energies of   Au  impurities  in ZB GaAs and  WZ GaAs are shown as dashed and solid lines, respectively. 
The thick lines represent stable charge states.
The vertical lines indicate the HOMO and LUMO levels in the defect-free supercell, same as Fig.~\ref{fig:EformVZBGaAs}. 
The left and right $y$-axes show the  formation energies under As-rich ($\mu_\mathrm{As}^{upper}$) and Ga-rich ($\mu_\mathrm{As}^{lower}$)
 conditions as described in the text.
\label{fig:EformAuZBGaAs}}
\end{figure}

We also study the energetics of substitutional Au impurities in GaAs crystals, and the 
computed  formation energies of Au$_{\rm As}$ and Au$_{\rm Ga}$ 
 as a function of  $\varepsilon_\mathrm{F}$ are shown in Fig.~\ref{fig:EformAuZBGaAs}.
The formation energies of  Au$_{\rm As}$ and Au$_{\rm Ga}$  are computed using Eq.~\ref{efmvatas}, with 
 the potential off-sets associated with Au$_\mathrm{As}$  and  Au$_{\rm Ga}$  listed in Table~\ref{vaupara} in the Appendix.
Moreover, the formation energies are specified for both  Ga-rich (right $y$-axis in Fig.~\ref{fig:EformAuZBGaAs})  
and As-rich (left $y$-axis Fig.~\ref{fig:EformAuZBGaAs}) conditions.
It is found that under an  As-rich growth condition an Au$_{\rm Ga}$ 
is energetically more favorable than an Au$_{\rm As}$ by about one to two  eV 
both in ZB GaAs and in WZ GaAs,
depending on the doping conditions. 
However, under a Ga-rich growth condition, the formation energy of an Au$_{\rm Ga}$ is 0.7~eV higher than that under an As-rich condition, 
while the formation energy of an Au$_{\rm As}$ becomes 0.7~eV lower. Therefore, under a Ga-rich condition, the Au$_{\rm As}$  defects 
turn out to be more stable  than Au$_{\rm Ga}$s. 
We expect that Au$_{\rm Ga}$ defects with  charge states from $2+$ to $2-$ may be stable under certain doping conditions. An Au$_{\rm Ga}$ has its charge transfer levels mostly in the lower part of the band gap, while the charge transfer levels of an Au$_{\rm As}$ lie at somewhat higher energies. The Au$_{\rm As}$ impurities occur preferentially as 2+, + or neutral defects, while the levels of negatively charged defects lie  probably already above the conduction band minimum of the host material.
Figure~\ref{fig:EformAuZBGaAs}(a) indicates the charge transfer levels of Au$_{\rm As}$s are deep, 
$E_v^\mathrm{ZB}+0.78$ eV and  $E_v^\mathrm{ZB}+0.91$ eV for ZB and WZ crystals, respectively. 
In contrast, an Au$_{\rm Ga}$ impurity  switches from a neutral state to a $1-$ charged state at 
$E_v^\mathrm{ZB}+0.22$  eV and $E_v^\mathrm{ZB}+0.35$  eV in ZB and WZ crystals,  respectively, as seen in Fig.~\ref{fig:EformAuZBGaAs}(b). 
As stated before, one should allow for a correction of about $~0.4$~eV to our $E_{v}$ value when comparing to experiment.
Experimental studies using deep-level transient spectroscopy and  
photoluminescence   spectroscopy~\cite{Hiesingerpssa76, Yanjes82, Pandianjjap91}
have identified an Au-related deep acceptor level of about 0.4 eV above the valence band in bulk GaAs.  
Moreover, we find an Au$_{\rm Ga}$ has a lower formation energy compared to an Au$_{\rm As}$ under the moderately As-rich growth conditions conventionally used. 
Taking into account 
the down-shift of $E_{v}$ with respect to the  DFT-GGA value (see e.g., Ref.~\onlinecite{Chen&Pasquarello2012}), 
it is plausible that 
the experimentally observed defect level at 0.4~eV above the valence band is indeed due to Au$_{\rm Ga}$ defects.

\begin{figure}
\includegraphics[width=7cm]{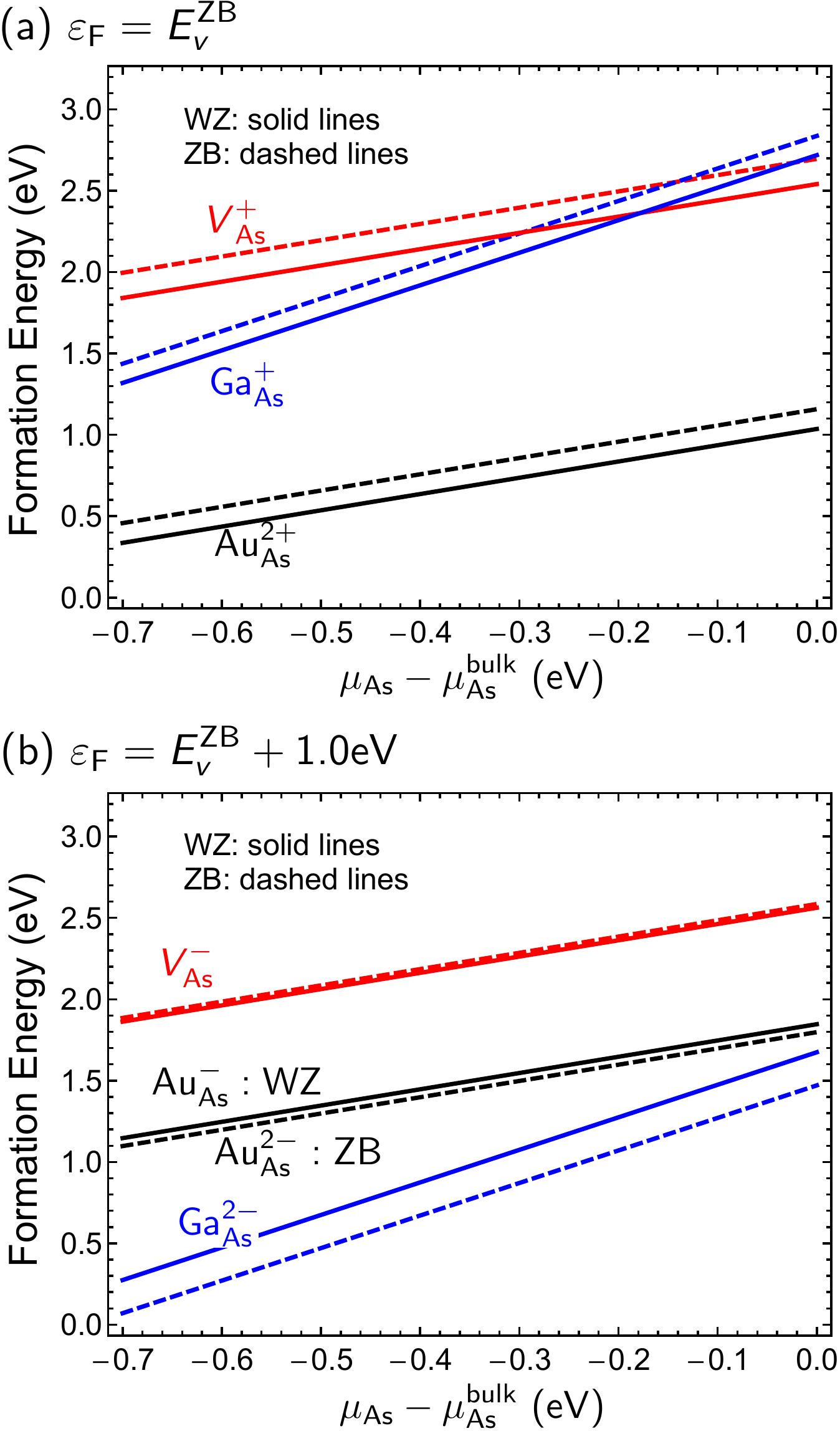}
\caption{(Color online)  The formation energies of $V_{\rm As}$,  Au$_{\rm As}$
and  Ga$_{\rm As}$ defects in GaAs crystals
as a function of  $\mu_{\rm As}$. 
In (a) and (b),  the Fermi energy is set to be at the calculated top of valence band and 1~eV above, respectively, to
describe a $p$-type or an $n$-type material.
Only the energetically  most favorable charge states are shown.
\label{fig:chempot}}
\end{figure}

We conclude this Section by discussing the energetics of defects caused by a deficiency of arsenic under varying growth conditions. 
In Fig.~\ref{fig:chempot}(a), we compare the formation energies of   the lowest energy $V_{\rm As}$,  Au$_{\rm As}$
and  Ga$_{\rm As}$ defects in $p$-type ZB and WZ GaAs crystals with the condition of $\varepsilon_{\rm F} - E_{v}^{\mathrm{ZB}} = 0$, as  a function of As chemical potential $\mu_{\rm As}$.
Our calculations show that  Au$_{\rm As}$ impurities have the lowest formation energies, while 
$V_{\rm As}$ and Ga$_{\rm As}$ 
defects have clearly higher formation energies. 
Figure~\ref{fig:chempot}(b) shows the same comparison for $n$-type ZB and WZ GaAs crystals 
with $\varepsilon_{\rm F} = E_{v}^{\mathrm{ZB}}+1.0$~eV.
Here, a Ga$_{\rm As}$ in the $2-$ charge state has the lowest formation energy, followed by Au$_{\rm As}$ and $V_{\rm As}$.
Consequently, when growing GaAs nanowires with a gold catalyst droplet, some Au$_{\rm As}$ defects will be
formed in the GaAs nanowire,
in particular for a $p$-doped material. 
The Au impurities act as deep centers and are thus detrimental to the optical properties of nanowires. 
In gold-free self-catalyzed growth,    
Ga$_{\rm As}$ and  $V_{\rm As}$ defects may exist 
in the nanowire, albeit at much smaller concentration than the Au defects 
if the material is $p$-doped.
Out of these three defect species, only the $V_{\rm As}$ defect is expected to be mobile via a hopping diffusion mechanism.  The mobility of the 
Ga$_{\rm As}$ and Au$_{\rm As}$ defects will most probably be vacancy-mediated, and thus require the $V_{\rm As}$ presence.  
Moreover, it is conceivable that $V_{\rm As}$ hopping diffusion could play a role in the nanowire growth process as a mechanism supplying arsenic to the interfacial growth zone between the nanowire tip and the catalyst droplet. Hence, we will investigate the detailed diffusion processes of  $V_{\rm As}$s
 in GaAs crystals in the following Section.

\section{A\lowercase{s} vacancy Diffusion in G\lowercase{a}A\lowercase{s}   \label{VGaAs}}

Arsenic vacancies in GaAs have been characterized experimentally 
by positron annihilation~\cite{SaarinenPRB44} and  by scanning-tunneling microscopy~\cite{GebauerPRB63}. 
For ZB GaAs, a number of theoretical studies, using (semi-)local DFT~\cite{PuskaPRB92, CarPRL94,Schulz} or hybrid functionals~\cite{Komsa,Komsa2012},
have been carried out, and the diffusion pathway of the arsenic vacancy has been investigated~\cite{Mousseau2007apa}. 
Interestingly, these calculations claimed the existence of a metastable interstitial state along the diffusion path in ZB GaAs, 
while this issue is unexplored for WZ GaAs.

\begin{figure}
\includegraphics[width=7cm]{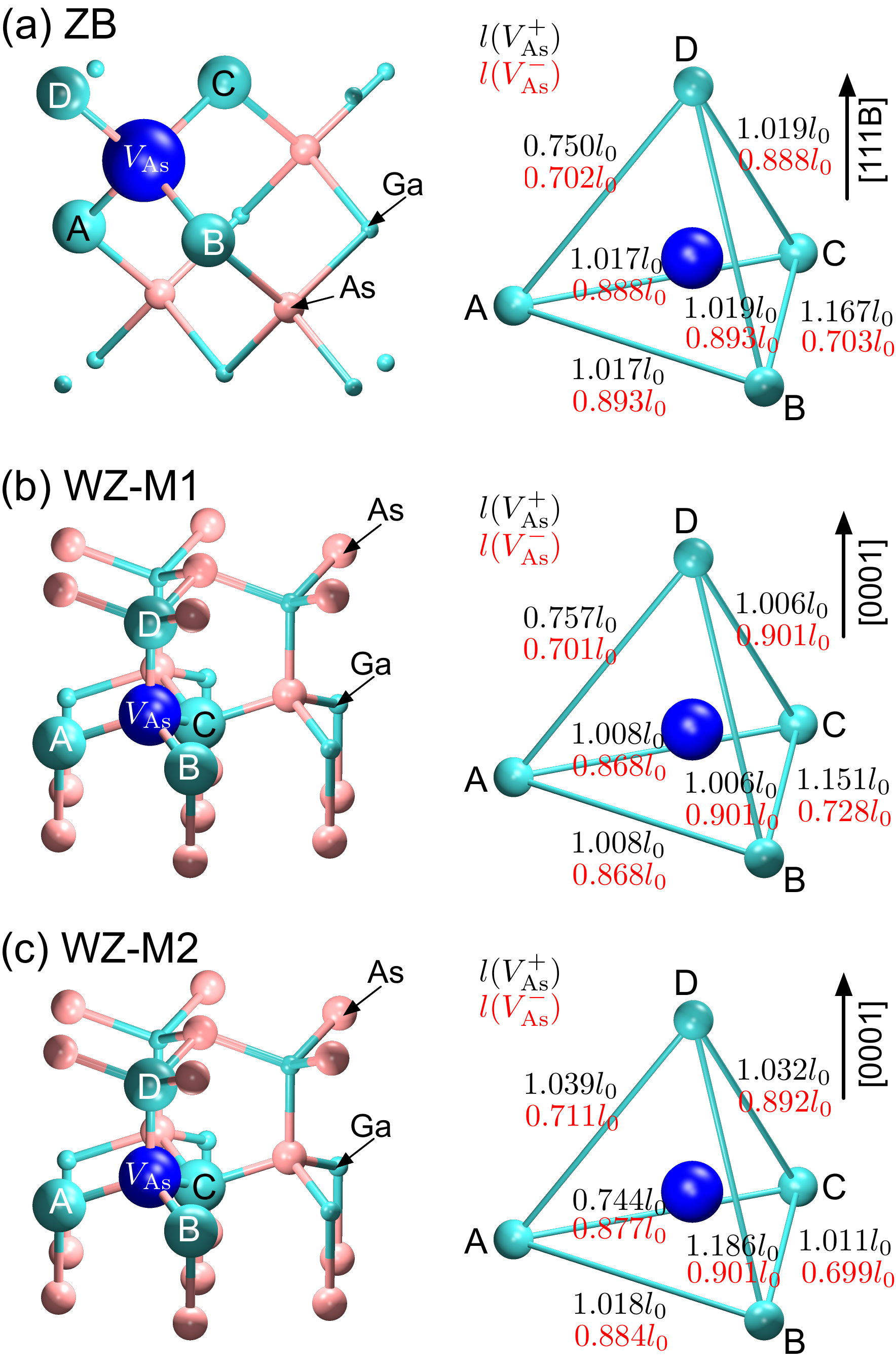}
\caption{(Color online)  The geometry of  $V_{\rm As}$s in  GaAs crystals. The relaxations of $V_{\rm As}^{+}$
and  $V_{\rm As}^{-}$ in ZB GaAs are illustrated in (a), while the  relaxations of two sets of 
stable $V_{\rm As}^{+}$ and  $V_{\rm As}^{-}$ in WZ GaAs are characterized in  (b) and (c), respectively.
We have found two stable $V_{\rm As}^{+}$s (M1 and M2) and  two  $V_{\rm As}^{-}$s (M1 and M2) in WZ GaAs.
 A $V_{\rm As}$  is  highlighted by a big blue sphere, and its four  neighboring Ga atoms are highlighted 
by  A, B, C, and D labeled spheres. These neighboring Ga atoms  form a tetrahedron.
The other attached As and Ga atoms are represented by small pink
red and large green spheres, respectively. The right panel indicates the relaxed edge length of the Ga
tetrahedron.
The upper and lower numbers refer to a $V_{\rm As}^+$ and a $V_{\rm As}^-$, respectively.
\label{fig:ZBtetra}}
\end{figure}

We start by characterizing the geometry of $V_{\rm As}$s in GaAs crystals in more detail. As shown in Fig.~\ref{fig:ZBtetra},
the neighboring Ga atoms with respect to a $V_{\rm As}$ define a Ga tetrahedron. 
To specify the local deformation of a crystal due to the  $V_{\rm As}$ presence, 
we study the relaxation of the edge length $l$ of the tetrahedron  
and compare it to the equilibrium distance between two  neighboring Ga atoms in the ZB (WZ) GaAs crystal of  $l_0$ = 4.07 (4.05)~\AA.
Similarly, the relaxed   tetrahedron volume $V$ will be 
compared to  the  equilibrium  tetrahedron volume in ZB (WZ) GaAs crystal of $V_0$ = 7.97 (7.93) \AA$^3$.

\begin{table}
\begin{tabular}{||l|c|c|c||}
\hline \hline
              & $V_\mathrm{As}^+$& $V_\mathrm{As}^0$& $V_\mathrm{As}^-$\\
\hline
ZB         & 0.922$V_0$& 0.757$V_0$ & 0.516$V_0$\\ \hline
WZ-M1 & 0.896$V_0$& 0.696$V_0$ & 0.518$V_0$\\  \hline
WZ-M2 &0.926$V_0$ &   & 0.512$V_0$\\  
\hline \hline
\end{tabular}
\caption{Ga tetrahedron volumes of $V_{\rm As}$s  in various charge states. $V_0$ represents the equilibrium tetrahedron volume in
a  ZB (WZ) GaAs crystal of $V_0 = 7.97 (7.93)$ \AA$^3$.}
\label{tab:Volume}
\end{table}

For a $V_{\rm As}^{-}$ in ZB GaAs, the edge lengths
of the Ga-atom tetrahedron are all contracted (see the right panel of
Fig.~\ref{fig:ZBtetra}(a)), resulting in a significantly reduced volume of
$V=0.516 V_0$ (48.4\%\ compression) as listed in Table~\ref{tab:Volume}.
Especially, the A-D and
B-C edges are most  contracted up to $l=0.702l_0$ (29.8~\%
compression), while other edges are compressed by $\sim$11~\%. The
deformation reduces the symmetry of the distorted Ga tetrahedron to $D_{2d}$ which
is in accordance with previous studies.~\cite{PuskaPRB92, CarPRL94} 
In WZ GaAs, there exist two different, but
energetically very similar structures (M1 and M2) for a relaxed $V_{\rm As}^{-}$ (see
Fig.~\ref{fig:ZBtetra}(b) and (c)). The WZ-M2 configuration is slightly
more stable than the WZ-M1 configuration in terms of energetics.  
Both WZ-M1 and WZ-M2  $V_{\rm As}^{-}$  configurations
exhibit  similar deformations as  a $V_{\rm As}^{-}$ in  ZB, 
and the relaxed tetrahedron volumes  are reduced by 48.4 \% to 48.8 \%, respectively. This is  comparable to
the ZB case.  The relaxed Ga tetrahedron for  WZ-M1 and WZ-M2  $V_{\rm As}^{-}$s possesses a  $C_{2v}$ and  a $C_{1}$
symmetry, respectively.

For  a $V_{\rm As}^{+}$, the tetrahedron of
neighboring Ga atoms contracts in volume by  only 7.8~\%\ in ZB
GaAs, whereas it contracts by 7.4 \% to 11.4 \% in WZ GaAs, as listed in Table~\ref{tab:Volume}.
In particular, $V_{\rm As}^{+}$s have  one edge  
compressed, while the other edges are slightly expanded. 
The WZ-M2  $V_{\rm As}^{+}$ configuration has the A-C edge  (within the  $ab$ plane) compressed.
 In contrast,  the WZ-M1  $V_{\rm As}^{+}$ configuration has the A-D edge 
(out of the  $ab$ plane) compressed.
In other words, the deformation of 
$V_{\rm As}^{+}$   
in WZ GaAs is anisotropic, and hence 
the relaxed Ga tetrahedron for  WZ-M1 and WZ-M2  $V_{\rm As}^{+}$s possesses $C_{2v}$ or $C_{1}$
symmetry, respectively. In addition, despite the neutral
state being unstable, we  mention that  the volumes of neutral As vacancies are
$0.757V_0$ in ZB and $0.696V_0$ in WZ-M1, respectively. It is worth noting  that the vacancy volume becomes smaller 
with increasing electronic charge. The physical reason for this general trend lies in the fact that the additional electrons in the vacancy allow for the filling of orbitals with bonding character that are formed by symmetry-adapted linear combinations of the Ga dangling-bond orbitals.

\begin{figure}
\includegraphics[width=8cm]{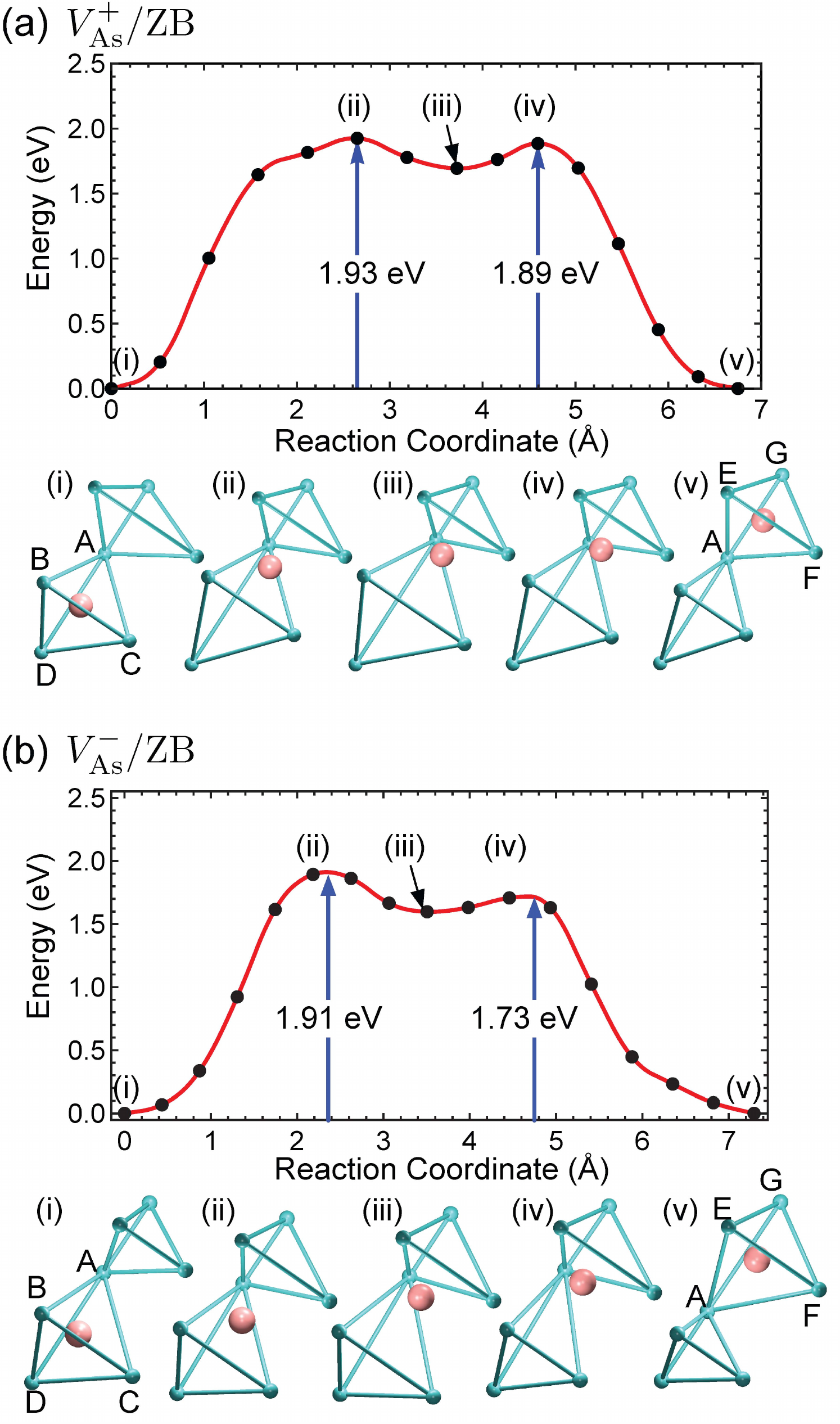}
\caption{(Color online)  The energy path of   $V_{\rm As}^{1+}$ (a) and  $V_{\rm As}^{1-}$ (b)  diffusion in  ZB GaAs. 
The structures of two ground state minima ($i$ and $v$), an interstitial  state ($iii$), and
 transition states ($ii$ and $iv$) are illustrated.
The As and Ga atoms are represented by lager pink  and small green spheres, respectively.  
\label{fig:ZBAsVacP1}}
\end{figure}

In the following, we assume that 
the  $V_{\rm As}$ diffusion proceeds via a neighboring As atom of the  arsenic sublattice hopping into the vacancy.
Schematically, an As atom  hops out of an initial 
Ga-tetrahedron cage into an adjacent target Ga-tetrahedron cage that previously enclosed the As vacancy site.
We refer to  the tetrahedron that surrounds the migrating As atom as the initial tetrahedron, and 
refer to the adjacent Ga-tetrahedron  that surrounds the target As vacancy site as the target tetrahedron.
We note that the reverse process occurs with the same probability, with the role of initial and target tetrahedron interchanged.
The diffusion process can be decomposed into an As escape from the initial tetrahedron, 
its interstitial motion between the two tetrahedra, and an entry  into the target tetrahedron. 
Note that one of the Ga atoms is located at the common apex of both  tetrahedra.

Since ZB GaAs is a cubic crystal, the four possible diffusion paths of a $V_{\rm As}$ to four neighboring
As sites are symmetrically equivalent.
The diffusion path shown Fig.~\ref{fig:ZBAsVacP1}  is sufficient to describe As diffusion in ZB GaAs.
 Figure~\ref{fig:ZBAsVacP1}(a) illustrates the detailed  diffusion process of  a $V_{\rm As}^{+}$ in 
ZB GaAs, indicating an overall migration barrier of  1.93~eV.
The local minimum configurations ($i$ and $v$ in Fig.~\ref{fig:ZBAsVacP1}(a)) as well as the saddle point 
configurations  ($ii$ and $iv$ in Fig.~\ref{fig:ZBAsVacP1}(a))  along the diffusion path are 
illustrated by the ball-and-stick models in the figure. 
In the target tetrahedron, the AG edge corresponds to the short edge (the AD in Fig.~\ref{fig:ZBtetra}(a) ).
By expanding the bond between As and Ga at the D-site, the diffusing  As atom escapes from the initial tetrahedron.
 During the escape process, the initial tetrahedron is expanding, while the target tetrahedron is contracting.
  After passing through the ABC face of the tetrahedron, the diffusing As atom arrives at 
 the first saddle point ($ii$) which represents the transition state for 
 $V_{\rm As}^{+}$ migration. In this configuration, the As-Ga bonds to
 the initial tetrahedron apexes  are broken except for the A-site. 
 The target tetrahedron volume is compressed from $0.922V_0$ to $0.719V_0$ which
 corresponds to the volume of a  neutral As vacancy (see Table~\ref{tab:Volume}).   
This observation implies that a positively charged As vacancy in the target tetrahedron 
attracts  electrons ahead of the  As migration.
When migrating toward the target tetrahedron in the interstitial space, the diffusing
 As atom reaches  an  interstitial  state ($iii$)  with a shallow energy dip.
At this moment, the diffusing As atom is located at an octahedral interstitial site in ZB GaAs.
 In this interstitial state, the migrating As creates two additional  bonds to target
 tetrahedron apexes, indicating that  the As atom starts to supply electrons to the
 target tetrahedron.  Consequently, the  target tetrahedron volume  further contracts to $0.662V_0$.
 This volume is still a bit larger than for a negatively
 charged As vacancy, but notably smaller than for a neutral As  vacancy.
Passing through the AEF face of the target tetrahedron,  the diffusing As atom enters the target vacancy.
The diffusion process is thus completed.
  When the As atom diffuses  into  the
 target tetrahedron, it establishes the As-Ga bond to the G-site. 
The migration from the octahedral interstitial site to the final state has a shallow migration barrier 
of 0.19~eV which is 0.04~eV lower than the backward diffusion barrier.
 We note that the contraction of the  AG edge of the empty target tetrahedron has moved over to the
 BC edge of the (finally empty) initial tetrahedron.

Figure~\ref{fig:ZBAsVacP1}(b) shows the  $V_{\rm As}^{-}$  diffusion 
path in ZB GaAs, which  is similar to $V_{\rm As}^{+}$ diffusion with 
a slightly lower migration barrier of 1.91~eV.
Again, we find the escape barrier from the initial
tetrahedron to be the  rate-limiting step.
In the interstitial space, there
is also an interstitial As configuration located in a shallow dip, and it occupies
an  octahedral interstitial site in ZB GaAs. The associated  
initial and  target tetrahedra 
at this stage 
have volumes of $1.102 V_0$ and  $0.673V_0$, respectively. 
We note that the tetrahedron volume 
expands while
loosing an electron (see Table~\ref{tab:Volume}). 
The diffusing As atom at the interstitial site has to overcome an entry barrier of 0.13 eV to migrate into the
target tetrahedron, which is 0.19 eV lower than that of the reverse diffusion process.
The  initial tetrahedron volume is reduced to the $V_{\rm As}^{-}$ value
after the diffusing As passes through  the second saddle point. 
We note that the two
saddle points along the diffusion path have  different energies. 
Although an oversimplified consideration solely based on the atomic positions suggests that a symmetric mechanism might exist, the 
observed symmetry breaking is physical 
because the charge balance between the two tetrahedra 
under the presence of the migrating As atom is non-symmetric. Therefore, 
it is meaningful to distinguish between 
the escape (first) saddle point  or the entry (second) saddle point to be rate-limiting.

Our results qualitatively agree with the previous theoretical study ~\cite{Mousseau2007apa}, which reported
migration barriers of 2.41 eV and 2.38 eV for $V_{\rm As}^{+}$ and
$V_{\rm As}^{-}$ diffusion in ZB GaAs, respectively, using atomic orbital basis sets. 
Both in their and in our work, the  computed  diffusion barrier of a $V_{\rm As}^{-}$ is slightly lower than that of a $V_{\rm As}^{+}$.
However, in our  plane-wave-based calculations, diffusion barriers of $V_{\rm
  As}^{+}$ and $V_{\rm As}^{-}$ are  computed to be 1.93 eV and 1.91 eV, respectively,
which are about 0.5 eV lower in energy than the values reported in
Ref.~\onlinecite{Mousseau2007apa}.   
We believe that these differences
are due to the more complete basis set for the wave functions in our
calculations, which allows for a more accurate description of the
energetics, in particular at transition state geometries. 

\begin{figure}
\includegraphics[width=8cm]{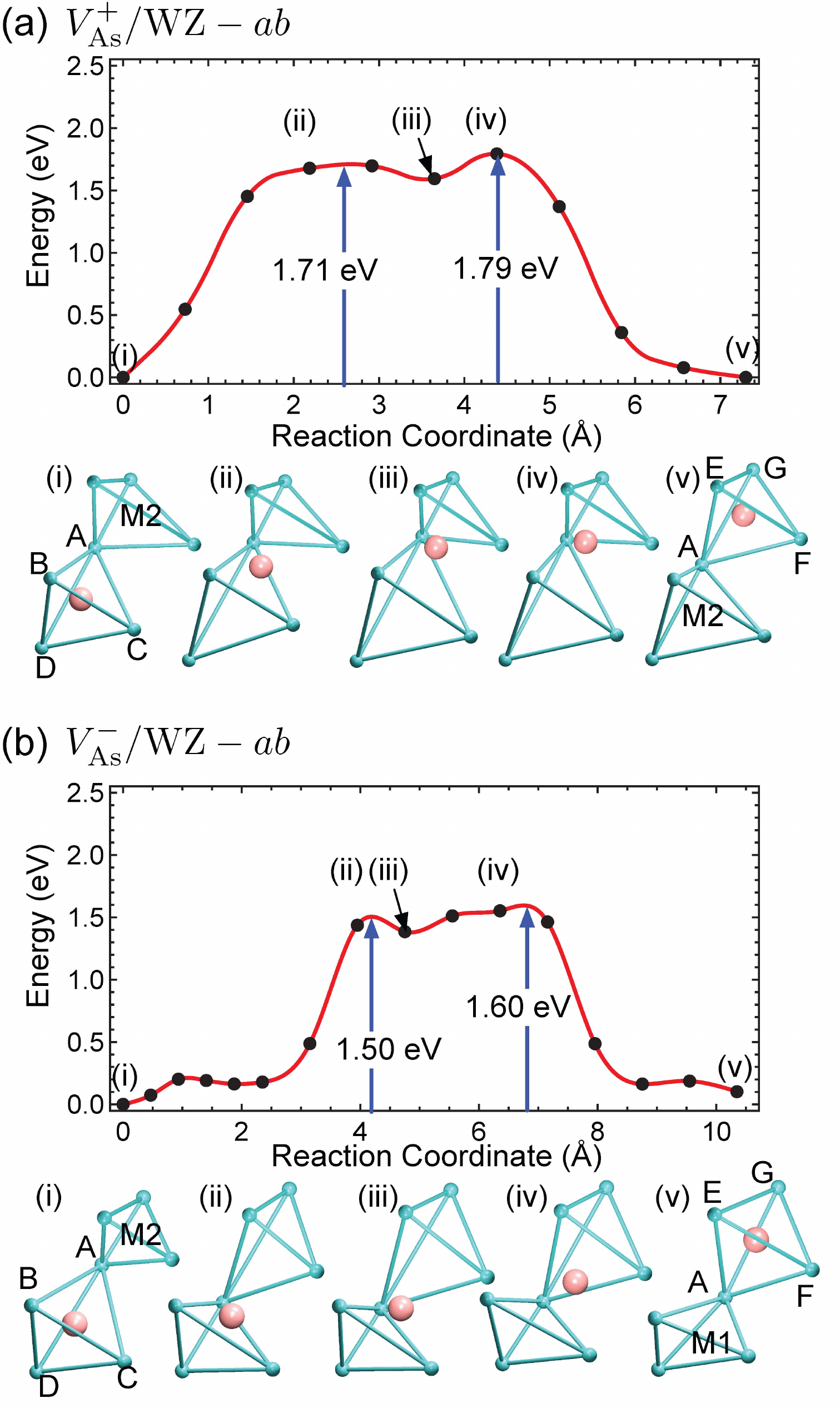}
\caption{(Color online)  The energy path of   $V_{\rm As}^{1+}$ (a) and  $V_{\rm As}^{1-}$ (b) 
 diffusion in the $ab$ plane of WZ GaAs.  
The structures of two ground state minima ($i$ and $v$), an interstitial  state ($iii$), and
 transition states ($ii$ and $iv$) are illustrated.
The color scheme is the same as in Fig.~\ref{fig:ZBAsVacP1}.
\label{fig:WzAsVac95Img9P1}}
\end{figure}

Next, we perform 
analogous  calculations for  $V_{\rm As}^{+}$ and $V_{\rm As}^{-} $ diffusion in WZ GaAs. 
Due to the lower crystal symmetry of WZ, one needs to consider  two pathways; i.e., an As vacancy can diffuse within the $ab$ plane by hopping from site 1 to site 2   
or along the $c$ axis  by hopping from site 
2 to site 3,  as illustrated in Fig.~\ref{fig:crystals}(b).
First, we discuss the vacancy diffusion within the  $ab$ plane. 
 Figure~\ref{fig:WzAsVac95Img9P1} shows that  $V_{\rm As}^{+}$ and $V_{\rm As}^{-}$ can diffuse 
 in the $ab$ plane in WZ GaAs with  migration barriers of 1.79 eV and 1.60 eV, respectively.
As discussed above,  there are two different minimum configurations for  $V_{\rm As}^+$ and  $V_{\rm As}^-$ defects, 
M1 and M2, in WZ GaAs. 
The transition from the more stable M2 to the M1 configuration costs relatively low activation energies, 
0.11~eV for a  $V_{\rm As}^{+}$ and 0.17~eV for a  $V_{\rm As}^{-}$. 
We basically find  similar diffusion processes as  in ZB GaAs, i.e.,
an As escape from the initial tetrahedron, an As motion within the interstitial space, and 
an As entry into the target tetrahedron. 
In the intermediate configuration in the $ab$ plane, the  As atom is   located in an octahedral interstitial site in WZ GaAs. 
Thus, the local symmetry of the intermediate is the same for WZ and ZB along this path.
However, the overall migration barriers are up to 0.2-0.3~eV lower in WZ than in ZB. 
In the $V_{\rm As}^{-}$ diffusion, we note that the volume contraction of the initial tetrahedron takes place earlier 
than in the other cases (see Fig.~\ref{fig:WzAsVac95Img9P1}(b)), i.e., its volume 
is reduced to $0.691V_0$ when As reaches the interstitial configuration.
In comparison, for  $V_{\rm As}^{-}$ diffusion in ZB GaAs, the  initial tetrahedron has a volume of $1.102V_0$ when the diffusing As is at the interstitial site.
Moreover, for $V_{\rm As}^{+}$s in WZ and ZB  crystals, the associated initial 
tetrahedron volume is about $1.8V_0$ to $1.9 V_0$ when the diffusing As is at the interstitial site.
This early compression of the initial tetrahedron for the 
 $V_{\rm As}^{-}$ diffusion also results in  a compressed interstitial space. 
In the interstitial space,  the diffusing  As atom forms three bonds to the target tetrahedron and one 
bond to the initial tetrahedron.  Hence, the  interstitial $V_{\rm As}^{-}$ is fully coordinated.
This could explain the  low migration barrier of 1.60 eV associated with the  
$V_{\rm As}^{-}$ diffusion within the $ab$ plane in  WZ GaAs.

\begin{figure}
\includegraphics[width=8cm]{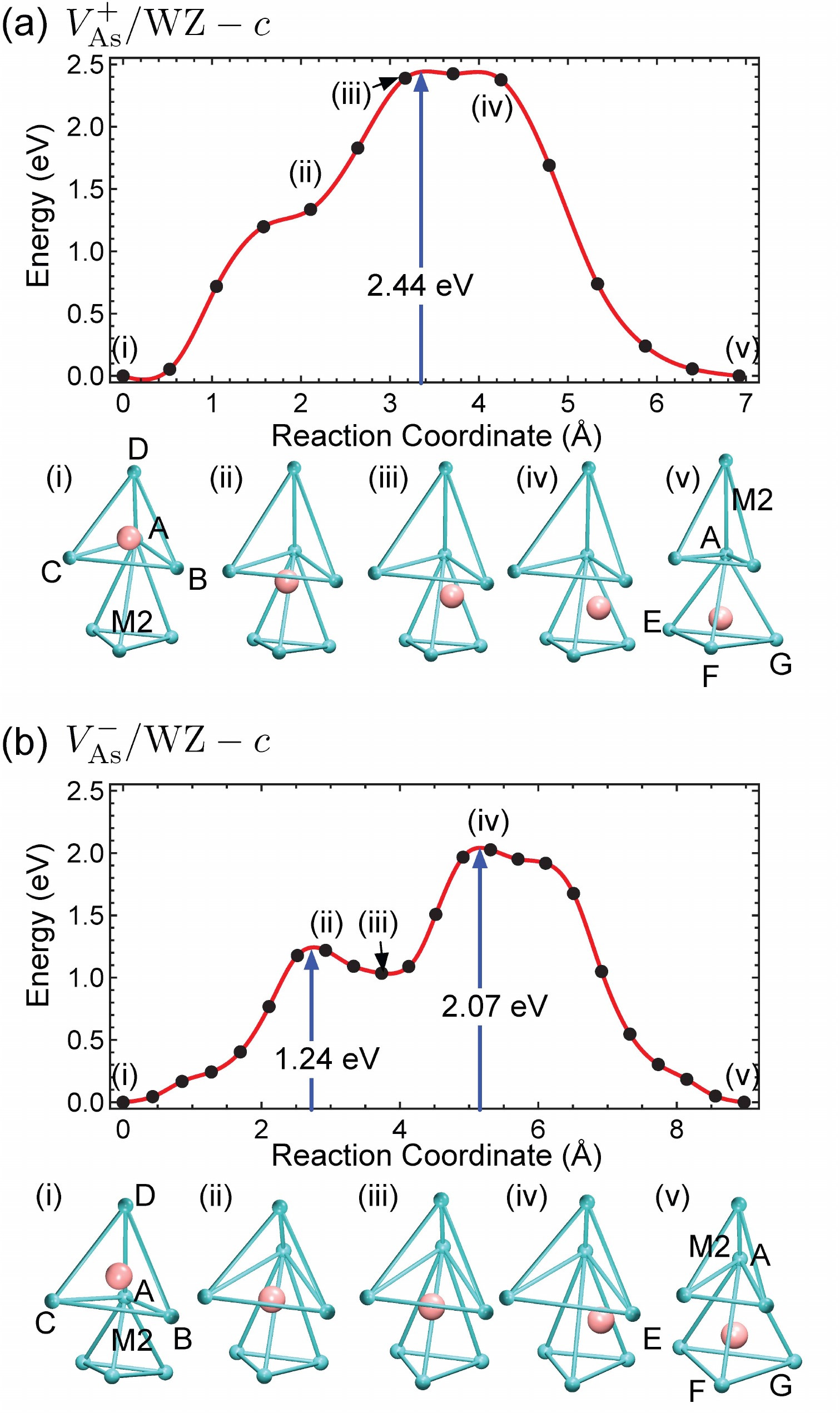}
\caption{(Color online) The energy path of   $V_{\rm As}^{+}$ (a) and  $V_{\rm As}^{-}$ (b) 
diffusion along the $c$-axis of in WZ GaAs.  
The structures of various stages ($i - v$) are illustrated.
The color scheme is the same as in Fig.~\ref{fig:ZBAsVacP1}.
\label{fig:WzAsVac95Img9P1-C}}
\end{figure}

Secondly, we discuss the $V_{\rm As}$ diffusion along the $c$-axis in WZ GaAs. 
Figure~\ref{fig:WzAsVac95Img9P1-C} shows that   $V_{\rm As}^{+}$ and $V_{\rm As}^{-}$ defects
can diffuse along the $c$-axis with relatively high migration barriers of 2.44~eV and 2.07~eV, respectively. 
Again, the diffusion process consists of the three steps introduced above:
The diffusing  As atom escapes from 
the initial  tetrahedron cage through the ABC face.
It enters the  neighboring octahedral interstitial area through the
 ABF face and finally hops into the target  tetrahedron through the AFG face.
When the  diffusing  As passes the ABF face, it almost crosses the AF edge. This may result in the 
high migration barriers associated with the $c$-axis  diffusion in WZ GaAs. 
For the  $V_{\rm As}^{-}$ diffusion, there exists 
a stable interstitial configuration ($iii$) as shown in Fig.~\ref{fig:WzAsVac95Img9P1-C}(b). For this configuration, the diffusing
As atom is located at a tetrahedral interstitial site.
It is  fully coordinated and
its energy is only 1.03~eV above  the ground state $i$ configuration. This may result in the lower $V_{\rm As}^{-}$  
migration barrier of 2.07 eV compared to the  $V_{\rm As}^{+}$ diffusion as shown in  Fig~\ref{fig:WzAsVac95Img9P1-C}.

\begin{table}
\caption{\label{barrier} Diffusion constants  for $V_{\rm As}$ diffusion in ZB and WZ GaAs at $T=700$ K.
All numbers are given in units of cm$^2$/s.
The   $V_{\rm As}$ diffusivity  along  the $c$ direction in WZ GaAs is the lowest.}
\begin{tabular}{||l|c|c|c||}
 \hline \hline
  & ZB & WZ-$ab$  &  WZ-$c$\\
  \hline
 $V_\mathrm{As}^+$ & $2.11\times10^{-16}$ & $2.13\times10^{-15}$ & $4.46\times10^{-20}$\\
 $V_\mathrm{As}^-$  & $2.94\times10^{-16}$ & $4.97\times10^{-14}$ & $2.05\times10^{-17}$\\
 \hline \hline
\end{tabular}
\end{table}

Finally, we discuss the possibility of reducing the concentration of $V_{\rm As}$ defects in nanowires by annealing under an arsenic atmosphere. We use the migration barrier of 1.93 eV  of a $V_{\rm As}^{+}$ in  ZB GaAs to estimate
 the required annealing time. The diffusivity associated
with the $V_{\rm As}$ can be estimated as
\begin{equation}
D \approx  a_0^2  \nu_{0}\exp\left(-\Delta E/k_{\rm B}T\right), 
\label{diff}
\end{equation}
where $\nu_{0} = 10^{13}$/s is the estimated attempt frequency, $k_{\rm B}$ is the Boltzmann constant, the annealing  temperature is set
to be $T=700$ K,  $\Delta E$  is the migration energy of an arsenic vacancy in GaAs, and
$a_0$  is  the distance between two neighboring As atoms in a GaAs crystal and is also the hop distance 
for $V_{\rm As}$ diffusion.
In a nanowire growing along the $[\bar{1}\bar{1}\bar{1}]$ direction, $V_{\rm As}$s are expected to diffuse to the side wall. Using a typical
radius of a GaAs nanowire, $R =20$ nm (Ref.~\onlinecite{Persson2004}), the annealing time required to 
remove the  $V_{\rm As}$s can be computed as $t = R^2/D = 5.26$ hours.
In WZ GaAs, the reduction of the energy barrier by 0.14~eV to 1.79~eV reduces the annealing time by a factor of 10 to 31.3~minutes.
Thus, we conclude that the $V_{\rm As}$s introduced in a GaAs nanowire during As-deficient growth can 
be annealed within a reasonable time frame at a temperature of 700 K.
We hereby summarize the diffusivities of $V_{\rm As}^+$ and $V_{\rm As}^-$ defects in ZB and WZ GaAs crystals
for various diffusion processes in Table~\ref{barrier}.

\section{Discussion and Summary \label{dis}}

Using DFT calculations, we have characterized the $V_{\rm As}$,
Ga$_{\rm As}$, and Au defects in GaAs crystals that may be introduced
in GaAs nanowires grown with the help of metal droplets under 
As-deficient conditions at the interfacial growth zone. The
substitutional Ga$_{\rm As}$, Au$_{\rm As}$, and Au$_{\rm Ga}$ defects
exhibit similar formation energies  and defect levels in both ZB
and WZ GaAs crystals. Moreover, a $V_{\rm As}$ defect behaves as a 'negative
$U$'-system, switching from a $V_{\rm As}^{+}$ in a  $p$-doped material to
a $V_{\rm As}^{-}$ in an $n$-doped material, for both GaAs crystals.  
In case that a Au droplet is used a catalyst for nanowire growth,
the formation of substitutional Au defects is possible and  is
found to be energetically more favorable than  the formation of 
Ga$_{\rm As}$ or $V_{\rm As}$ defects.  Moreover, we have shown that
it is energetically more favorable by about 1 to 2~eV for an Au
substitutional defect to replace a lattice Ga atom than a lattice As
atom in either ZB or WZ GaAs.  
Given that DFT-GGA calculations give a too small band gap, 
the calculated acceptor level of Au$_{\rm Ga}$ 
at $E_{v} + 0.22$ eV is found to be in reasonable agreement with the
experimental value of  $E_{v} + 0.4$ eV in ZB GaAs.

In  Au-free, self-catalyzed growth of nanowires that are grown at higher
temperatures,~\cite{MartelliNT08} the formation of 
Ga$_{\rm As}$ could  be expected. In particular, for Ga-rich growth conditions and $n$-type material, the Ga$_{\rm As}^{2-}$ species has a low formation energy. This could lead to growth of non-stoichiometric, Ga-enriched GaAs.\cite{Persson2004, TambeNL10} The incorporation of Ga$_{\rm As}^{2-}$ will counteract the $n$-doping by deliberately added donor species. This may explain why it is rather difficult to obtain $n$-type conductivity in ZB GaAs nanowires, and only high concentrations of  Sn$_{\rm Ga}$ have proven successful so far.~\cite{Gutsche2011} 
Under less Ga-rich growth conditions, the formation of 
$V_{\rm As}$s could be expected, and
their diffusion might even play a role for the material transport in
the growth of a GaAs nanowire. 
We note that As interstitials, which could in principle also contribute to As mass transport, tend to have a higher formation energy than  $V_{\rm As}$s, apart from very As-rich conditions.~\cite{Schulz} 
Since we are interested in the As-deficient growth conditions below the nanoparticle, we don't consider the As interstitial diffusion here.
Our results show that i) $V_{\rm As}$s can
diffuse within ZB GaAs with a migration barrier of about 1.9 eV; and
ii) $V_{\rm As}$s diffuse favorably within the $ab$ plane in WZ
GaAs with somewhat lower migration barriers of 1.6 to 1.8 eV. 
Based on these
results, we estimate that it takes about five hours to anneal the
$V_{\rm As}$s at 700~K in a ZB GaAs nanowire, but only 30~minutes in a WZ
nanowire.  Thus, an annealing of GaAs nanowires under an arsenic
atmosphere could be useful to obtain samples with a longer lifetime for
the charge carriers.
However, the diffusivity of $V_{\rm As}$s in the nanowires is too low to contribute substantially to the arsenic supply at the nanowire growth zone, in particular in WZ GaAs, where the diffusion barrier along the $c$-axis is higher than in ZB GaAs.
 
\section*{Acknowledgments}
We acknowledge Center for Computational Sciences and Simulation (CCSS) of University Duisburg-Essen for the 
computer time and the Deutsche Forschungsgemeinschaft DFG for the financial support through the project KR 2057/5-1. 

\appendix*
\section{ Parameters for computing formation energies}

The Madelung
constants, the calculated dielectric constants, and  the  dimensions  of ZB and WZ GaAs
crystals are listed in Table~\ref{vpara}. 
$L$ represents the
distance between the nearest neighbor  defects within a supercell. It is taken to be length
of the $a$-axis of a 216 atom ZB and a 96 atom WZ GaAs crystal as listed in Table~\ref{vpara}.
The dielectric constants are the  $\epsilon_{aa}$ components of pertinent dielectric tensors.
They include both electronic and ionic contributions. 
The values of the potential
off-set corrections for various defect and impurity calculations are
listed in Table~\ref{vaupara}.

\begin{table}
\caption{\label{vpara} The  Madelung constants ($\alpha$),
the dielectric constants  ($\epsilon$)  and  the  dimension of GaAs supercells  ($L$) for computing the formation energy 
as described in Eq.~\ref{efmvatas}. The values are  listed for both ZB and WZ GaAs supercells.
For convenience, $\epsilon$ is given  in units of $e$\AA$^{-1}$V$^{-1}$. 
The dimensionless dielectric constant is obtained by multiplying this value with 14.4.
}

\begin{tabular}{||c|c|c|c||} \hline \hline
   &   $\alpha$ & $\epsilon$ &  $L$ (\AA) \\ \hline \hline
ZB  &   1.638     &   1.18   &  17.29 \\ 
WZ  &    1.641    &   1.12   &  12.16 \\
\hline \hline
\end{tabular}
\end{table}

\begin{table}
\caption{\label{vaupara} The values of the potential off-set corrections for 
  $V_{\rm As}$ ($\Delta V_{V {\rm  @ As }}$),   Au$_{\rm As}$ ($\Delta V_{\rm Au @ As }$),
  Au$_{\rm Ga}$  ($\Delta V_{\rm Au @ Ga }$), and Ga$_{\rm As}$ ($\Delta V_{\rm Ga @ As }$) defects in WZ and ZB GaAs crystals.
All numbers are given in units of eV.}
\begin{tabular}{||c|c|c|c|c||} \hline \hline
   &  $\Delta V_{V {\rm  @ As}}$   &$\Delta V_{\rm Au @ As}$  &  $\Delta V_{\rm Au @ Ga }$  &  $\Delta V_{\rm Ga @ As }$  \\ \hline \hline
ZB &  $-0.05$   & $-0.01$          &   $-0.04$ & $-0.05$  \\ 
WZ &  $-0.09$   &$-0.02$          &   $-0.02$ & $-0.09$  \\
\hline \hline
\end{tabular}
\end{table}


\end{document}